\documentclass[final,twocolumn,times,5p]{elsarticle}
\usepackage{amsfonts}
\usepackage{amsmath}
\usepackage{mathtools}

\usepackage{multirow}
\usepackage{diagbox}
\usepackage{booktabs}
\usepackage{algorithm}
\usepackage{algorithmic}
\usepackage{graphicx}
\usepackage{enumitem}
\usepackage{float}
\usepackage{stfloats}
\usepackage{makecell}
\usepackage[font=small,skip=0pt]{caption}
\usepackage{hyperref}
\usepackage{url}
\graphicspath{ {./images/} }

\journal{Computers \& Security}

\begin{document}

\begin{frontmatter}

\title{Few-shot Weakly-supervised Cybersecurity Anomaly Detection}
%
%

\author{Rahul Kale, and Vrizlynn L. L. Thing}

\address{Singapore Technologies Engineering, Singapore}

\begin{abstract}
With increased reliance on Internet based technologies, cyberattacks compromising users' sensitive data are becoming more prevalent. The scale and frequency of these attacks are escalating rapidly, affecting systems and devices connected to the Internet. The traditional defense mechanisms may not be sufficiently equipped to handle the complex and ever-changing new threats. The significant breakthroughs in the machine learning methods including deep learning, had attracted interests from the cybersecurity research community for further enhancements in the existing anomaly detection methods. Unfortunately, collecting labelled anomaly data for all new evolving and sophisticated attacks is not practical. Training and tuning the machine learning model for anomaly detection using only a handful of labelled data samples is a pragmatic approach. Therefore, few-shot weakly supervised anomaly detection is an encouraging research direction. In this paper, we propose an enhancement to an existing few-shot weakly-supervised deep learning anomaly detection framework. This framework incorporates data augmentation, representation learning and ordinal regression. We then evaluated and showed the performance of our implemented framework on three benchmark datasets: NSL-KDD, CIC-IDS2018, and TON\_IoT.
\end{abstract}


\end{frontmatter}

\section{Introduction}\label{sec:introduction}

With the rapid rise in the networking technologies, multiple industries such as manufacturing \cite{giehl2019framework}, healthcare\cite{sethuraman2020cyber}, financial institutions\cite{vedral2021vulnerability} and even worldwide government agencies \cite{covid} are adopting internet based solutions for services and data storage. These systems have to manage, process or store sensitive user data to provide the services and fulfill their functionalities. Therefore, cyberattackers are targeting the Internet based devices to compromise and gain unauthorized access to the sensitive data. The end users of these systems may not be equipped with the necessary knowledge or technical tools to protect themselves from these growing threats. Hence, intrusion detection becomes an integral part of the initial defense against the cyberattacks.

Anomaly detection serves as an important tool for the intrusion detection systems \cite{lazarevic2003comparative}. Many machine learning based techniques are proposed in literature for anomaly detection or intrusion detection. These techniques can be broadly classified into categories such as fully-supervised learning \cite{naseer2018enhanced,zhong2020helad,imtiaz, du2021network}, semi-supervised learning \cite{akcay2018ganomaly, ruff2019deep,vercruyssen2018semi}, unsupervised learning \cite{chen2021daemon,merill}, hybrid learning \cite{ahmad2019hybrid, Garg2019} or weakly-supervised anomaly detection  \cite{sheynin2021hierarchical, pang2021explainable, ding2021few, zhou2020siamese}. Supervised Learning methods learn the patterns of entire training data based on the labels provided whereas semi-supervised learning methods learn the patterns of only the normal data from the training set to identify anomaly data. Unsupervised learning methods identify the anomalies by learning the patterns in the unlabelled data. The input data does not contain any label information in this case.

Firstly, it is quite challenging to collect large amount of data for all the available threats. Secondly, it is expensive to obtain labelled attack samples as it may need manual labelling of the threats. Furthermore, with the fully labelled data approach, it is not possible to incorporate defense against new or future threats. Therefore, label information dependent supervised and semi-supervised machine learning methods may not be the best overall choice for anomaly detection. Even though it is generally difficult to obtain large-scale labelled dataset for supervised anomaly detection, obtaining a few labelled anomaly samples is comparatively less expensive and more practical. Unsupervised methods do not utilize any labelled data, even if a few such labelled samples are available. Hence, data-efficient anomaly detection \cite{pang2021deep} using few-shot weakly supervised methods which utilize the limited number of labelled anomaly samples, is an interesting research direction explored in recent literature.

Few-shot learning (FSL) is a type of machine learning method where the training dataset contains limited information. Few-shot learning techniques such as zero-shot learning enable machine learning models to separate two classes that are not even present in the training data. The main idea here is to associate known and unknown classes through supplementary information which encodes the differentiating characteristics of the samples \cite{xian2017zero}.
Few-shot learning has important use cases when it is challenging to find sufficient training data (e.g. rare medical cases) or the data labelling process is extremely expensive.
The name weakly-supervised/Few-shot method is to suggest that the available labelled training sample set is small compared to usual supervised method. It also signifies that the small training set is not complete i.e. it does not contain information/samples from all the available anomaly classes\cite{pangweak}. In addition, the anomalies grouped under single category may not have similar features e.g. features of samples from DDoS attack subcategory may not be similar. Therefore, it is important for the weakly-supervised method to be able to identify anomalies based on the knowledge of very few known class anomalies.
As the amount of training data required by FSL model is comparatively lower, training data annotation and collection costs are minimized. Furthermore, computational processing requirements are also reduced for lower amount of training data.

In this work, we are exploring an enhancement to the weakly-supervised anomaly detection network proposed in \cite{pangweak}. The overall framework consists of three stages: The first stage is data augmentation, which is generating additional data samples through the efficient use of the available limited set of labelled data samples. The enhanced or augmented dataset is used as the training dataset for the second stage of the framework. The purpose of the second stage is to learn the compressed representation of the enhanced dataset without compromising the characteristics of the original individual samples. Anomaly scores are generated by learning the combined compressed representation of the enhanced data samples. The final stage of the framework performs the function of calculating the loss through ordinal regression \cite{pangweak} during the training phase of the framework. The overall framework is based on the assumption that the anomaly samples are not statistically dominant in the given dataset \cite{foorthuis2021nature}. Additionally, the number of labelled anomaly samples available are extremely small compared to the number of unlabelled samples available in the dataset.

To summarize, we proposed an enhanced three-stage anomaly detection network based on few-shot weakly-supervised learning in this paper. We evaluate the performance of the proposed framework on NSL-KDD, CIC-IDS2018, and TON\_IoT datasets.
The rest of the paper is organized as follows: Related works in the recent literatures are reviewed in Section \ref{sec:related}. The details of the proposed framework are presented in Section \ref{sec:framework}. The details about the experimental setup and dataset are discussed in Section \ref{sec:setup}. The results are presented in Section \ref{sec:results}. The key points and findings of all the experiments are summarized in Section \ref{sec:discussion}. Finally, Section \ref{sec:conclusion} concludes the paper.

\section{Related Works}\label{sec:related}
In the recent literature, various methodologies are proposed for anomaly detection and intrusion detection problems in different domains.
For NSL-KDD dataset, Naseer et al. explored the suitability of multiple deep learning methods for the IDS based on anomaly detection in \cite{naseer2018enhanced}. Zhong et al. proposed a multi-stage HELAD algorithm in \cite{zhong2020helad} for network traffic anomaly detection. It uses a combination of damped incremental statistics algorithm for feature extraction, trained autoencoder for abnormal traffic identification, and trained LSTM to obtain the final network anomaly detection. For anomaly detection in IoT networks, Ullah and Mahmoud designed and developed a multi-class classification model based on CNN in \cite{imtiaz}. Du and Zhang \cite{du2021network} proposed a selective ensemble algorithm for network anomaly detection. By under-sampling the majority class samples, multiple training sets are generated for multiple sub-classifiers. The algorithm was designed to handle the imbalanced datasets inherently.

For semi-supervised learning, authors in \cite{akcay2018ganomaly} proposed deep anomaly detection based on a generative adversarial network where the network was trained on the normal samples and the anomaly samples were identified based on the reconstruction error of compressed representation of normal samples. Ruff et al. proposed a deep semi-supervised anomaly detection framework based on minimizing and maximizing the entropy of the latent distribution for normal and anomaly data, respectively in \cite{ruff2019deep}. Their approach was based on the Infomax principle. Authors in \cite{vercruyssen2018semi} proposed constraint-clustering-based anomaly detection framework through active-learning strategy for monitoring water consumption in supermarkets by using the labels as constraints for guiding the clustering.

Unsupervised algorithms were also explored for anomaly detection problems.
 Merril and Eskandarian \cite{merill} provide enhanced Autoencoder(AE) training for unsupervised anomaly detection. Cumulative error scoring, percentile loss, and knee detection for early stopping are the three modifications proposed to improve the performance over conventional AE.
The early network traffic anomaly detection method proposed by Hwang et al in \cite{hwang2020unsupervised}, adopts a combination of CNN and AE for profiling and filtering of anomalous traffic to provide high accuracy at extremely low false-positive rate.
Chen et al.\cite{chen2021daemon} propose robust unsupervised anomaly detection DAEMON framework for multivariate time series. It also provides anomaly interpretations through reconstruction error of the univariate time series, and robustness through the regularization of reconstructed data and hidden samples via adversarial method. Therefore, the unsupervised anomaly detection methods can be used for a variety of datasets.
In addition to unsupervised learning, hybrid anomaly detection methods are also proposed in the literature. Ahmad et al. proposed a hybrid anomaly detection method by utilizing K-medoid customized clustering technique for misdirection and blackhole attacks in wireless sensor networks in \cite{ahmad2019hybrid}. Authors in \cite{Garg2019} proposed a two-step hybrid model for cloud datacenter networks anomaly detection through the utilization of extended improved versions of Gray Wolf Optimization and Convolutional Neural Networks. 

Supervised learning based methods need large and exhaustive set of labelled data which is not practical. Unsupervised methods fail to benefit from the availability of handful of labelled anomaly samples.
According to a recent survey in 2021 \cite{pang2021deep}, few-shot weakly supervised deep learning methods form an encouraging direction to approach anomaly detection problems.
Authors in \cite{sheynin2021hierarchical} proposed a hierarchical generative model to identify the multi-scale patch distribution of training images and further augmented the representation through image transformation to calculate anomaly scores using patch-based vote aggregation. Pang et al. proposed a prior-driven anomaly detection framework to perform differentiable learning of anomaly scores instead of feature learning for the limited anomaly samples in \cite{pang2021explainable} and additionally provided anomaly explanation. A meta-learning cross-network approach for the graph deviation networks was proposed in \cite{ding2021few} so that meta-knowledge from multiple auxiliary networks can be transferred to these new enhanced graph neural networks. A neural deviation learning based approach was proposed by Pang et al. in \cite{devnet} which utilizes limited labelled anomaly samples and a reference score of normal samples along with a scoring network for the anomaly detection. Zhou et al. proposed a Siamese Convolutional Neural Network in \cite{zhou2020siamese} which was designed to calculate the distances of data samples based on their optimized representations with the purpose of addressing the over-fitting problem using the few-shot learning model. Authors in \cite{sultani2018real} propose a weakly-labeled method for anomaly detection in surveillance video data without the need of specific temporal information in the anomaly video samples. This approach is based on the multiple instance ranking which utilizes features extracted from both normal and anomaly video segments.

Hence, few-shot weakly supervised methods have shown great potential for anomaly detection in different domains. To incorporate the benefits of these machine learning methods, we are exploring their implementation in cybersecurity domain for our proposed method.

\section{Anomaly Detection Framework}\label{sec:framework}

\begin{figure}[htb]
	\centering
	\includegraphics[scale=0.25]{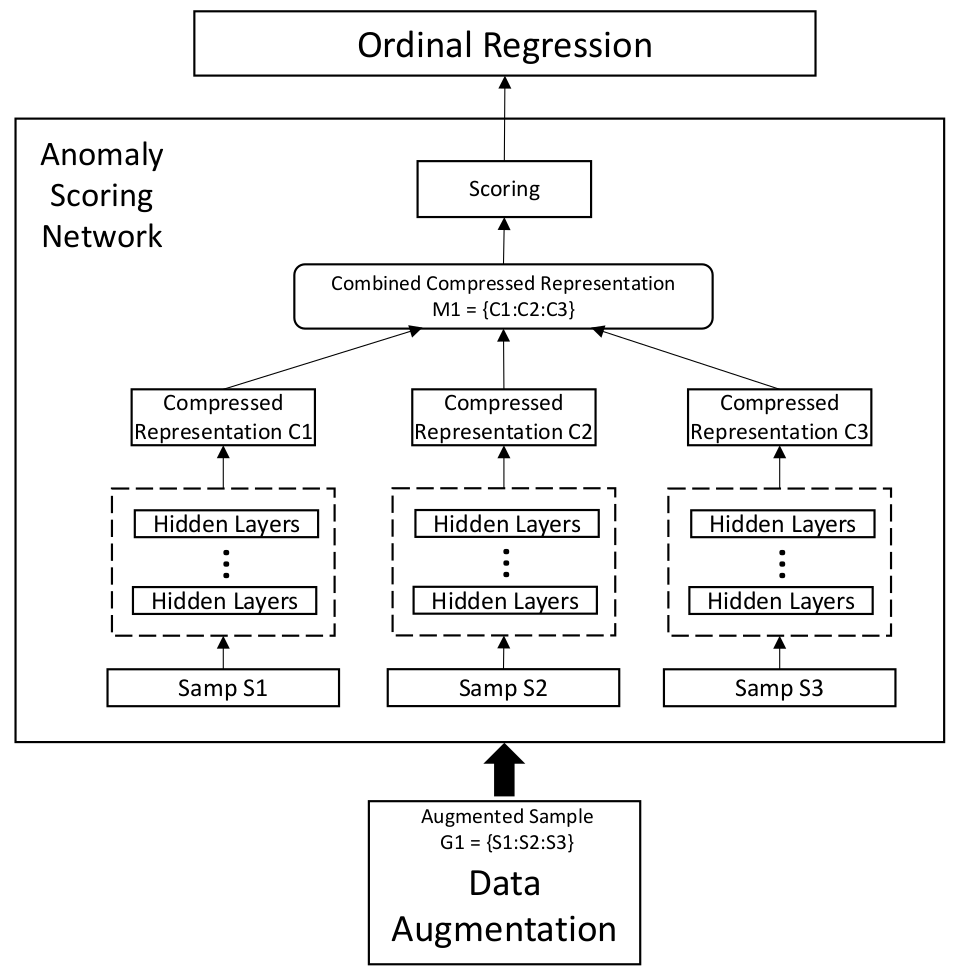}
	\caption{Anomaly Detection Framework}
	\label{fig:framework}
\end{figure}

The anomaly detection framework consists of three stages as shown in Fig. \ref{fig:framework}.

The main objective of the data augmentation part, is about generating additional data samples through the efficient use of the available limited set of labelled data samples. The data-preprocessing steps such as handling of missing data, categorical features and normalization of the data are performed in this stage of the framework. The processed data is passed through the data augmentation steps. This enhanced dataset serves as the training dataset for the second stage of the framework. In the second stage, anomaly scores are generated by learning the combined compressed representation of the enhanced data samples. The final part of the framework performs the function of calculating the loss through ordinal regression  during the training phase of the framework.

\subsection{Problem Formulation}
Let $\mathcal{X} = \{\mathbf{x}_{i}\}$, $i = 1,\dots,K$ , $\mathbf{x} \in \mathbb{R}^n$ be the entire dataset of the input data samples. This dataset comprises of two parts: first part is the unlabelled dataset $U$ containing $N$ samples and second part is $A$ number of labelled anomaly samples such that $A \ll N$. Also, the labelled anomaly samples do not cover all the anomaly classes. The objective of the anomaly detection framework is to generate anomaly scores $s_{i}$ for classification of data sample $\mathbf{x}_{i}$ given a threshold $AD_{th}$ such that:

\begin{equation*}
	y_{i} =
	\begin{cases}
		0, & \text{if $s_i < AD_{th}$}\\
		1, & \text{if $s_i \geq AD_{th}$}\\
	\end{cases}
\end{equation*}
\noindent where $y_i$ denotes the predicted label. $y_i = 0$ indicates that the sample $x_i$ is a normal sample, whereas $y_i = 1$ indicates that the sample $x_i$ is an anomaly. The details about the anomaly score $s_i$ are discussed in Section \ref{sec:ors}.

We will now describe the enhancements to the three stages of the anomaly detection framework compared to \cite{pangweak}.

\subsection{Stage 1: Pre-processing and Data Augmentation} \label{sec:aug}
\subsubsection{Pre-processing}
Before performing the data augmentation steps, some pre-processing is required for the input data. One of the pre-processing steps includes encoding of the categorical features within the dataset. For categorical variables with ranked ordering, ordinal encoding is used. For categorical variable containing finite number distinct values without any ranked ordering, one-hot encoding is used.

For normalization, we adopt Min-Max Normalization such that the normalized value $x^{norm}_{i,j}$ for sample $i$ and feature $j$ is mapped within the range [0,1] and it can be calculated as:

\begin{equation*}
	 x^{norm}_{i,j} = \frac{x_{i,j} - MIN(j)}{MAX(j) - MIN(j)}
\end{equation*}

\noindent where $x_{i,j}$ is the value of $j^{th}$ feature for the $i^{th}$ sample, $MIN(j)$ is the minimum value for the $j^{th}$ feature and $MAX(j)$ is the maximum value for the $j^{th}$ feature.

\subsubsection{Data Augmentation}

\begin{figure}[htb]
	\centering
	\includegraphics[width=3.2in]{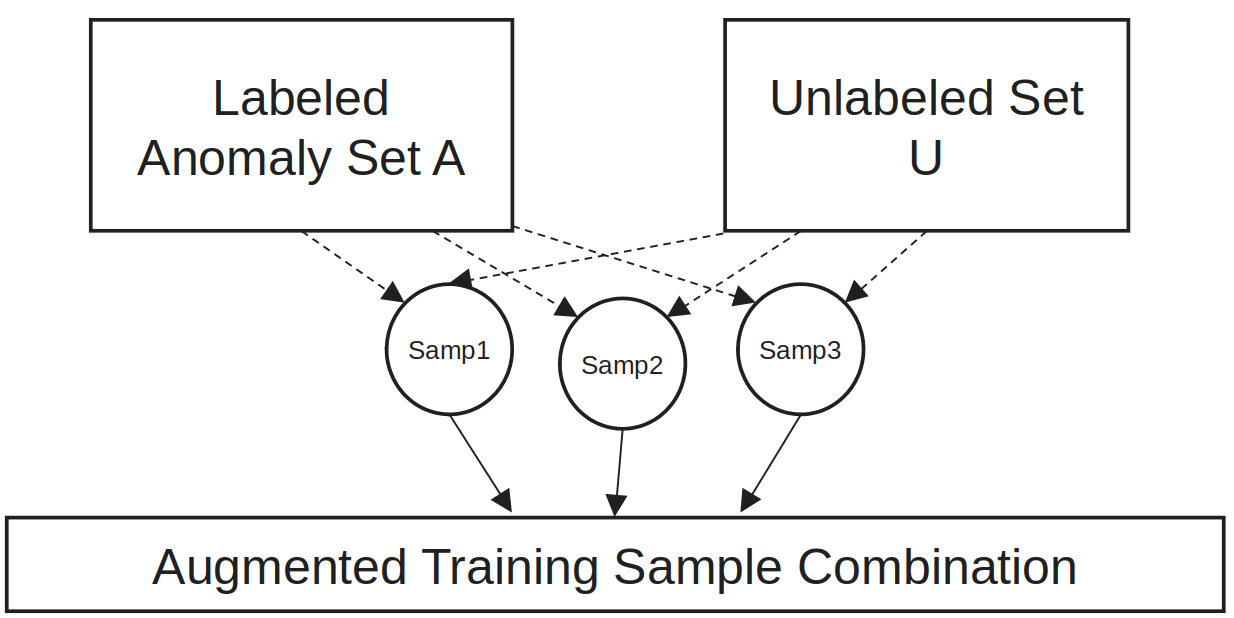}
	\caption{Data Augmented Triplet Sample}
	\label{fig:augsamp}
\end{figure}

The number of available labelled anomaly samples is quite limited. Therefore, it is extremely important to generate augmented dataset for training the network in the second part of the framework.
Samples are taken randomly from Labelled Anomaly Set $A$ and Unlabelled Set $U$ to create augmented training data of instance triplets as shown in Fig.\ref{fig:augsamp}.

Three samples will be randomly picked from the sets A and U. Each combination (class) will be given a distinct class label and some of the combinations are grouped under single class label as shown in Table \ref{table:augment_class}. Class labels are integers such that the ordinal regression can be performed in Stage 3. An important point to note here is that $C1 > C2 > C3 > C4 \geq 0$.

	\begin{table}[htbp]
	\centering
	\caption{ClassLabels for Augmented Data}
	\label{table:augment_class}
	{\begin{tabular}{|>{\centering}m{1.1cm}|>{\centering}m{1.1cm}|>{\centering}m{1.1cm}|>{\centering}m{1.4cm}|c|}
			\hline
			Sample 1 & Sample 2 & Sample 3 & Combin-ation & ClassLabel \\
			\hline
			A & A & A & C(A,A,A) & C1 \\
			\hline
			A & A & U & \multicolumn{1}{c|}{} & \multicolumn{1}{c|}{}\\
			\cline{1-3}
			A & U & A & C(A,A,U) & C2\\
			\cline{1-3}
			U & A & A & \multicolumn{1}{c|}{} & \multicolumn{1}{c|}{}\\
			\hline
			A & U & U & \multicolumn{1}{c|}{} & \multicolumn{1}{c|}{} \\
			\cline{1-3}
			U & A & U & C(A,U,U) & C3 \\
			\cline{1-3}
			U & U & A & \multicolumn{1}{c|}{} & \multicolumn{1}{c|}{}\\
			\hline
			U & U & U & C(U,U,U) & C4 \\
			\hline
			
	\end{tabular}}
\end{table}	

The rationale behind $C1>C2>C3>C4$ is related to the number of anomaly samples present in the corresponding augmented data sample. The anomaly scoring function is designed in such a way that the scores for the anomaly samples should be higher than the normal/unlabelled sample. The ordinal labels $C1$ to $C4$ are directly associated with the class combinations. The augmented combination $[C(A,A,A):C1]$ is formed using 3 anomaly samples,whereas $[C(A,A,U):C2]$ is formed using 2 anomaly samples. Similarly, $[C(A,U,U):C3]$ is formed using only 1 anomaly sample, whereas [C(U,U,U):C4] is formed using without any anomaly sample. As these ordinal labels are directly used for loss function, it is essential to maintain the higher anomaly score for augmented samples with more anomaly samples, which dictates $C1>C2>C3>C4 \geq 0$.

The main motivation for considering 3-sample/triplet combination is related to better data augmentation combinations over the 2-sample method proposed in \cite{pangweak}. In 2-sample method, the possible combinations are $(A,U), (A,A)$ and $(U,U)$. By including an additional labelled sample, the limited labelled data further augments and generalizes the new training set from available dataset. Consider known anomaly classes $A1, A2, A3$ present in the labelled anomaly set. For two sample method, the number of possible combinations are $3^2$. In case of 3 sample method, the number of possible combinations are $3^3$. This way, the network has more useful combinations to learn the representation of $C(A,A,A)$ combination for anomaly identification and scoring. This difference in number of combinations will be even higher when the included number of known classes is more.

It is important to consider the range and values of $C1$, $C2$, $C3$, and $C4$.
As established earlier, $C1>C2>C3>C4 \geq 0$. So, now consider,  $ (C1 – C2) = (C2 – C3) = (C3 – C4) = m$ where $m$ is the common difference. If value of $m$ is too small, there would not be sufficient distinction among the class labels. If the value of $m$ is too large, it may not fit for all the classes properly. 
Authors in \cite{pangweak} varied the value between 1 to 10 to identify the optimal value with respect to different test datasets. They identified that $m=4$ provided generally the best results with performance decreasing for higher values. Therefore, using the same consideration, we are fixing our values as $C1=12, C2=8, C3=4, C4=0$.

\subsection{Stage 2: Anomaly Scoring Network} \label{sec:asn}

\begin{figure}[htb]
	\centering
	\includegraphics[width=3in]{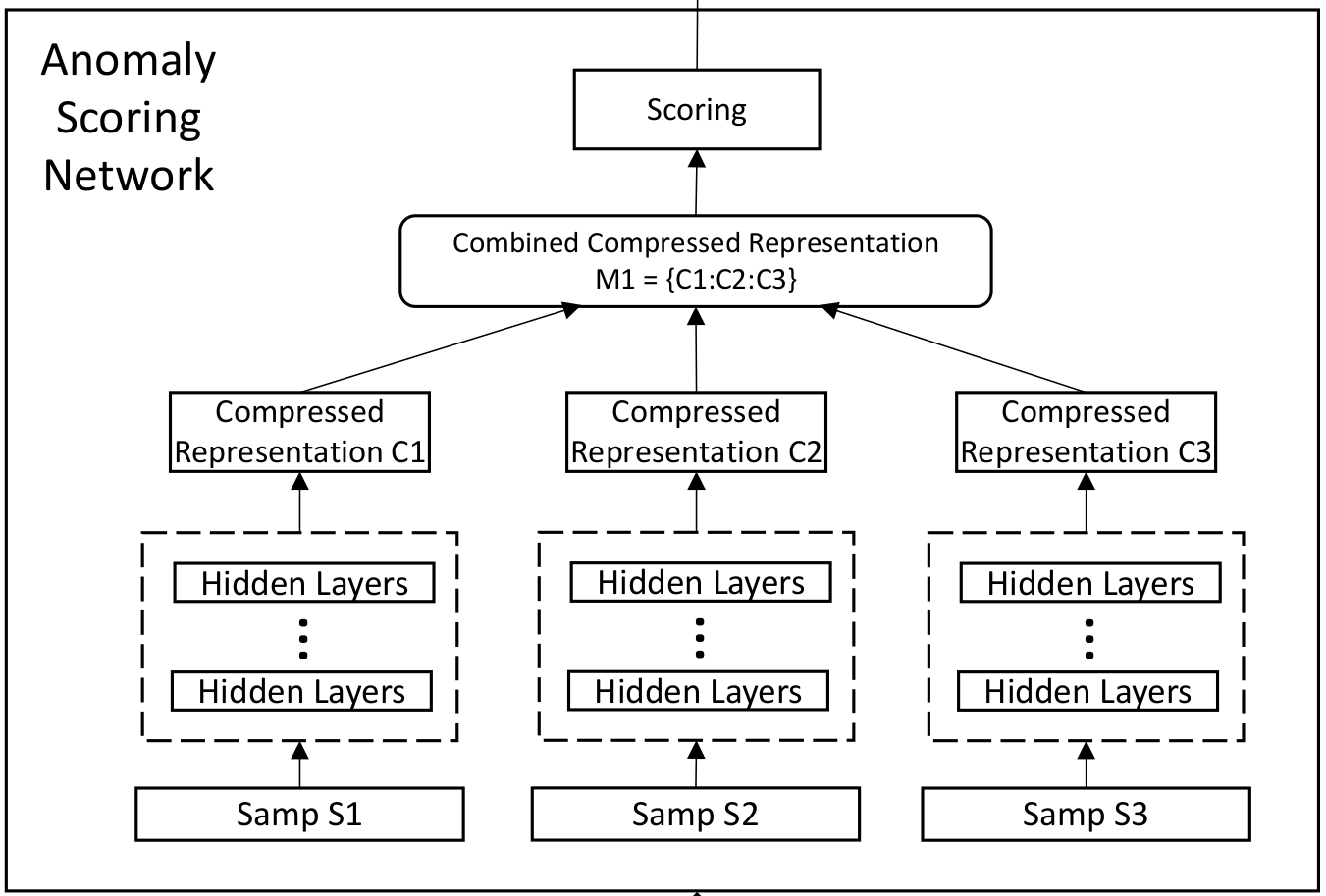}
	\caption{Anomaly Scoring Network}
	\label{fig:anomaly_scoring}
\end{figure}

The anomaly scoring network consists of hidden network structure in the second stage as shown in the Fig. \ref{fig:anomaly_scoring}. The triplets of networks used for calculating the hidden networks share the weight parameters. For our datasets, we have used multi-layer perceptrons in the hidden layers.

Though the overall structure is similar to \cite{pangweak}, there are three sub-networks instead of two, to accommodate for triplet augmented sample. The three sub-networks are structurally identical, and they share the weights. The weight sharing of the networks has two main benefits. First, each augmented sample is split in the anomaly scoring network in such a way that each individual network learns the representation of only a single sample from the original dataset. Hence, it is still optimized for original dataset rather than augmented sample space. Secondly, by sharing the weights within the networks, the three-stream networks do not need to be individually optimized. Thus, it reduces the overall computational complexity.

Consider an augmented sample $G1$, created from three data samples $S1, S2$ \& $S3$, is given as an input to the anomaly scoring stage shown in Fig.\ref{fig:anomaly_scoring}. The input dimensions of each sub-network are adjusted such that it will process only a single sample from the original dataset individually. These sub-networks first reduce the dimensions of the input data by mapping it to a lower dimensional representation. Let's consider the output of each sub-network as compressed representations $C1, C2$ \& $C3$ for samples $S1, S2$ \& $S3$, respectively. The combined compressed representation $M1$ is formed by combining the three lower dimensional representations $C1, C2$ \& $C3$. This combined  representation is used by the linear unit to compute the anomaly scores in the \textit{Scoring} block from Fig.\ref{fig:anomaly_scoring}. The entire network is mapping the inputs to scalar anomaly scores and it is end-to-end trainable\cite{pangweak}.

\subsection{Stage 3: Ordinal Regression and Scoring Phase} \label{sec:ors}

Ordinal regression will be performed in the third stage of the framework similar to \cite{pangweak}. Absolute prediction error loss is used to minimize the difference between the prediction scores and the ordinal labels. The loss function $L$ will be as per Eq.\ref{eq:lossfunc}:
\begin{multline}
	L(\phi((x_i,x_j,x_k);\theta),y_{\{x_i,x_j,x_k\}}) \\
	= |y_{\{x_i,x_j,x_k\}} - \phi((x_i,x_j,x_k);\theta)|
	\label{eq:lossfunc}
\end{multline}
\noindent where $(x_i,x_j,x_k)$ refer to the augmented sample formed using the combination of three input samples $x_i,x_j \& x_k$; $\theta$ refers to the corresponding weight matrices; $\phi$ refers to the overall anomaly scoring network; $y_{}$ refers to the ordinal labels given to the augmented samples.

The loss function for the proposed method is similar to the loss function of 2-sample method \cite{pangweak}. However, the anomaly score in proposed method is obtained using 3 samples. The objective function is similar as well. But again, the key difference is in the anomaly score which is calculated using 3 samples. The overall objective function is given in Eq.\ref{eq:objectivefunc}:
\begin{multline}
	\resizebox{0.45\textwidth}{!}{$\underset{\theta}{argmin}\frac{1}{|B|} \sum_{(x_i,x_j,x_k,y_{\{x_i,x_j,x_k\}})\in B}|y_{\{x_i,x_j,x_k\}} - \phi((x_i,x_j,x_k);\theta)|	+ \lambda R(\theta)$}
	\label{eq:objectivefunc}
\end{multline}
\noindent where $(x_i,x_j,x_k)$ refer to the augmented sample formed using the combination of three input samples $x_i,x_j \& x_k$; $\theta$ refers to the corresponding weight matrices; $\phi$ refers to the overall anomaly scoring network; $y_{}$ refers to the ordinal labels given to the augmented samples; $B$ refers to the batch size;
$\lambda$ is the hyperparameter for the regularization $R()$.

The behavior of third stage will be different during the test phase. During the test phase, for each test sample $T$, the following steps will be performed:

\begin{itemize}
\item[1]Select two reference samples from labelled anomaly set $A$ at random to form first augmented sample $C(A,A,T)$.
\item[2]Calculate the anomaly score $S1$ for $C(A,A,T)$
\item[3]Select two reference samples from unlabelled set $U$ at random will be chosen to form second augmented sample $C(T,U,U)$.
\item[4]Calculate the anomaly score $S2$ for $C(T,U,U)$
\item[5]Record the combination of $S1$ and $S2$ as $S3$
\item[6]Repeat Steps 1 to 5 for 30 times. Each time, the samples from $A$ and unlabelled set $U$ will be picked randomly.
\item[7]The average of S3 over 30 runs will determine if the sample $T$ is an anomaly
\end{itemize}

If the test sample T is an Anomaly, the first augmented sample is of the form $C(A,A,A)$ and the second augmented sample is of the form $C(A,U,U)$. Therefore, $S1$ will be closer to the ordinal label $C1$ and $S2$ will be closer to the value $C3$.
If the test sample T is normal, the first augmented sample is of the form $C(A,A,U)$ and the second augmented sample is of the form $C(U,U,U)$. Therefore, $S1$ will be closer to the ordinal label $C2$ and $S2$ will be closer to the value $C4$. Therefore, $(C1+C3) > (C2+C4)$ will identify $T$ as an anomaly. As established earlier, $C1>C2>C3>C4 \geq 0$. That is, if the sample is an anomaly, the combined score will be higher.

\section{Experimental Setup}\label{sec:setup}
To evaluate the performance of the proposed framework, we will use the three datasets: NSL-KDD \cite{nsl}, CIC-IDS2018\cite{ids2018} and TON\_IoT(Win10) \cite{moustafa4}. We will now briefly describe each dataset.

\begin{itemize}[leftmargin=*]
	\item \textbf{NSL-KDD}\cite{nsl}:
	NSL-KDD dataset was developed with the aim of overcoming the inherent limitations  from the KDD99 dataset mentioned in \cite{tavallaee2009detailed}. Some of the advantages of NSL-KDD over KDD99 include reduced redundant training data; removal of duplicate records from test data; and sufficient number of records in both training and testing dataset. NSL-KDD dataset contains 43 features. The categorical features such as "protocol\_type", "service" and "flag" are one-hot encoded during the pre-processing step. The attack categories for the NSL-KDD dataset are primarily classified into 4 main types: probe, Remote to Local (R2L), User to Root (U2R), and Denial of Service (DoS).
	These four main attack types will be considered as a single group of "Attack" category samples.
	The sample distribution details of NSL-KDD dataset is given in Table\ref{table:main_samp_distrib}.

	\item \textbf{CIC-IDS2018}\cite{ids2018}:	
	CIC-IDS2018 dataset was developed to create a versatile intrusion detection benchmark dataset. It utilizes detailed user-profiles which contain comprehensive representations of the various network activities and intrusion characteristics. The network traffic was collected using CICFlowMeter-V3\cite{ids2018} by capturing the system logs of multiple machines to generate CIC-IDS2018 dataset with 80 features. The features such as timestamp that would not influence the anomaly detection are dropped from the dataset during the pre-processing. The CIC-IDS2018 includes 7 attack types.
	The extracted sample distribution details of CIC-IDS2018 dataset is given in Table\ref{table:main_samp_distrib}.

	\item \textbf{TON\_IOT(Windows10)}\cite{moustafa4}:	
	The TON\_IoT dataset includes the telemetry data collected from IoT and IIoT sensors, Linux and Windows Operating Systems and other network traffic data captured in practical large-scale networks. We are using the Windows 10 dataset in this paper, which was captured using the Performance Monitor Tool on Windows 10 systems to include common operating system activities related to general functions such as network and memory activities.	
	The dataset contains 124 features and 7 attack types. The features such as timestamp that would not influence the anomaly detection are dropped from the dataset during the pre-processing.
	The sample distribution details of source TON\_IoT dataset is given in Table\ref{table:main_samp_distrib}.
\end{itemize}

	\begin{table}[htbp]
			\centering
			\caption{Sample Distribution for source datasets}
			\label{table:main_samp_distrib}
			{\begin{tabular}{|c|>{\centering}m{1.5cm}|p{1.6cm}|}
					\hline
					Dataset & Normal Samples & Anomalous Samples \\
					\hline
					NSL-KDD & 76432 & 66684 \\
					\hline
					CIC-IDS2018 & 2856035 & 1669364 \\
					\hline
					TON\_IoT (Win10) & 10000 & 11104 \\
					\hline
					
			\end{tabular}}
		\end{table}

Before we describe the experimental design, we would like to highlight two considerations. Firstly, even though the three datasets considered in this paper contain the label information, it is not utilized during the training of the algorithm and it is dropped from the unlabelled training dataset. The label information is used only to create the small labelled anomaly set. Secondly, for the evaluation of the anomaly detection, the different attack types are considered as a single group of "Attack" category samples.

To evaluate the effectiveness of the algorithms for anomaly detection, the experiment was performed for each of the three datasets independently and the same experimental procedure was followed for all three datasets. For the purpose of explaining the experimental steps, consider a placeholder dataset $D$ which represents one of the three main datasets: NSL-KDD, CIC-IDS2018 and TON\_IoT. The steps can be described in the following manner.:
\begin{itemize}[leftmargin=*]
	\item From Dataset $D$, 5 random subsets will be created. These subsets are non-overlapping and do not contain any common samples. We will refer to each subset as a SampleSet. Each SampleSet is formed by picking the normal data samples and anomaly data samples completely randomly from the main dataset D. There is no overlap between the normal or anomalous samples among the SampleSets.
	
	The size of each SampleSet is identical with same number of normal data and anomaly data samples. In each SampleSet, the anomaly percentage is maintained at around 10\%.
	
	\item The final result is calculated by taking the mean of results obtained for each SampleSet.
\end{itemize}

The size of each SampleSet for the three datasets is given in Table \ref{tab:sampset_numbers}.

\begin{table*}[htbp]
\centering
\caption{SampleSet Distribution}
\label{tab:sampset_numbers}
		{\begin{tabular}{|>{\centering}m{2.2cm}|>{\centering}m{2.1cm}|>{\centering}m{2.4cm}|>{\centering}m{2.2cm}|>{\centering}m{3.3cm}|>{\raggedright\arraybackslash}m{2.5cm}|}
				\hline
				Dataset & Normal Samples (N) & Anomaly Samples (A\_total) & Labelled Anomaly (A\_labelled) & AvailableDataset D \\ (N + A\_total - A\_labelled) & Anomaly\% in AvailableDataset D \\
				\hline
				NSL-KDD SampleSet & 13460 & 1620 & 120 & 14960 & 10.03 \\
				\hline
				CIC-IDS2018 SampleSet & 20000 & 2350 & 120 & 22230 & 10.03 \\
				\hline
				TON\_IoT (Win10) SampleSet & 1948 & 337 & 120 & 2165 & 10.02 \\
				\hline
				
		\end{tabular}}
	\end{table*}

Independent experiments are performed for each SampleSet in the proposed framework. The testing procedure is as explained in Section \ref{sec:ors}.

\subsection{Evaluation Metrics}
For the evaluation of anomaly detection, several performance metrics such as accuracy, precision, recall, etc. are available. Because of the inherent imbalance between the number of anomaly samples and the number of normal data samples, the accuracy metric may not correctly represent the performance of the evaluating methods. We have adopted Receiver Operating Characteristics (ROC), True Positive Rate (TPR) and False Positive Rate (FPR) as the evaluation metrics. True Positives(TP), True Negatives(TN), False Positives(FP) and False Negatives (FN) are also provided for reference.

\begin{equation}
	TPR = \frac{TP}{TP + FN};
	FPR = \frac{FP}{FP + TN}
\end{equation}

\noindent where TP denotes True Positives or the number of correctly identified attack samples; FN denotes False Negatives or the number of attack samples incorrectly identified as normal samples; FP denotes False Positives or the number of incorrectly identified normal samples as attacks; and TN denotes the True Negatives or the number of correctly identified normal samples.
In addition, receiver operating characteristic curve (ROC curve)~\cite{roc} is used to evaluate the anomaly detection results.
ROC can be elucidated as the probability of assigning an anomalous sample a higher score compared to a normal instance \cite{davis2006relationship}.

\subsection{Comparison Methods}
The performance of the proposed algorithm will be compared against the weakly-supervised 2-sample method in \cite{pangweak}, Deviation Network algorithm \cite{devnet} and the Hybrid algorithm for anomaly detection \cite{rahulhybrid}. We will now briefly explain each algorithms.\\
\noindent \textbf{1. Weakly-supervised 2-sample:} The weakly-supervised method proposed in \cite{pangweak} forms the basis for the proposed framework. It consists of data augmentation, anomaly scoring and ordinal regression stages similar to the stages described in Section 3. This method uses instance pairs for data augmentation and hence requires only two sets of identical networks in anomaly scoring stage.\\
\noindent \textbf{2. Deviation Network:} The deviation network method proposed in \cite{devnet} is a deep anomaly-detection method comprising of an anomaly scoring network and a reference score generator which optimize the anomaly scores using the deviation loss function. This method also utilizes only a limited number of labelled anomaly samples during its training. The source code is provided in \cite{devnet}.\\
\noindent \textbf{3. Hybrid algorithm:} The hybrid framework consists of a combination of unsupervised, semi-supervised and supervised learning for intrusion detection. However, we will consider only the anomaly detection part of the hybrid framework for the comparison with the proposed framework. Therefore, this hybrid framework will function as an unsupervised method and will not require the limited labelled anomaly set.

The training and test set will be kept identical for each of the methods and the proposed framework to obtain fair comparison.

\subsection{System Setup}
The proposed framework is implemented entirely in Python. Keras with Tensorflow backend was utilized for the implementation. Performance evaluation experiments were conducted on a system with UbuntuOS 20.04, 16-core CPU with 64GB RAM and Nvidia RTX3070 GPU with 8GB VRAM.

\section{Results}\label{sec:results}

We will now evaluate the anomaly detection of the proposed framework through the area under the ROC curve (AUROC) and TPR-FPR comparison. The performance evaluation was carried out across three experiments. For the first experiment, both the number of labelled anomalies and the anomaly percentage were kept fixed. For second experiment, the number of labelled anomalies was varied while keeping the anomaly percentage fixed. In the third experiment, the number of labelled anomalies was fixed whereas the anomaly percentage was varied.

\subsection{Experiment 1: AUROC results with the labelled anomaly set}

For this experiment, the number of labelled anomalies and the anomaly percentage in training dataset was kept fixed at $60$ and $10$ respectively.

For each of the evaluation methods, the mean of the AUROC results across different SampleSets for three datasets are summarized in Table \ref{table:exp1auroc}.
	\begin{table}
	\centering
	\caption{Experiment 1: AUROC results (mean$\pm$stddev)}
	\label{table:exp1auroc}
			{\begin{tabular}{|>{\centering}m{1.3cm}|>{\centering}m{1.4cm}|>{\centering}m{1.2cm}|>{\centering}m{1.2cm}|p{1.1cm}|}
					\hline
					\textbf{Dataset} & {\makecell{\textbf{Weakly}\\\textbf{2-sample}}} & \textbf{DevNet} &\textbf{Hybrid} & \textbf{Propo-sed}\\
					\hline
					NSL-KDD & $0.94\pm0.01$ & $0.97\pm0.006$ & $0.91\pm0.01$ & $0.94\pm0.01$\\ 
					\hline
					CIC-IDS2018 & $0.75\pm0.016$ & $0.80\pm0.014$ & $0.69\pm0.012$ & $0.76\pm0.015$\\ 
					\hline
					TON\_IoT (Win10) & $0.99\pm0.01$ & $0.99\pm0.01$ & $0.90\pm0.02$ & $0.98\pm0.01$ \\ 
					\hline
			\end{tabular}}
		\end{table}
	
It can be observed from the results, that the AUROC of the proposed framework is either on par or better than 2-sample weakly supervised method. Here, the Deviation Network outperforms the other algorithms for all three datasets. The weakly supervised 2-sample method was designed with the assumption that the anomaly percentage in the given dataset will be low \cite{pangweak}. As the proposed method is enhanced version of that method, the same assumption applies, however, including additional sample for the augmented dataset has resulted in better performance in some cases.

\subsection{Experiment 2: Varying the number of labelled anomalies}
To evaluate the impact of number of labelled anomalies on the overall anomaly detection, we varied the number of anomalies while keeping the anomaly percentage fixed at 10\%. For this experiment, the number of labelled anomalies were varied between $[30,60,120]$ for each dataset. As the hybrid method does not require the labelled anomalies, we have provided a single datapoint for hybrid method as a reference for this experiment.

For all three datasets, generally the AUROC performance improved with the increase in the number of labelled anomalies as shown in Tables \ref{table:exp2resnsl} to \ref{table:exp2cmton}.

For NSL-KDD dataset, as shown in Tables \ref{table:exp2resnsl} and \ref{table:exp2cmnsl}, the AUROC performance for the proposed method improved when the number of labelled anomalies was increased from 30 to 60. Additional anomaly samples may have provided further support to data augmentation for training the anomaly scoring network.


\begin{table*}[!bp]
	\centering
	\caption{Experiment 2 Results (NSL-KDD): AUROC, TPR, FPR}
	\label{table:exp2resnsl}
			{\begin{tabular}{|c|c|c|c|c|c|c|c|c|c|c|c|c|}%
					\cline{1-13}
					\multicolumn{1}{|c|}{\textbf{\# of}} & \multicolumn{12}{|c|}{Method} \\
					\cline{2-13}
					\multicolumn{1}{|c|}{\textbf{labelled}} & \multicolumn{3}{|c|}{\textbf{Weakly 2-sample}} & \multicolumn{3}{|c|}{\textbf{DevNet}} & \multicolumn{3}{|c|}{\textbf{Hybrid}} & \multicolumn{3}{|c|}{\textbf{Proposed}} \\%
					\cline{2-13}
					\textbf{anomalies} & {\makecell{{AU-}\\{ROC}}} & TPR & FPR & {\makecell{{AU-}\\{ROC}}} & TPR & FPR & {\makecell{{AU-}\\{ROC}}} & TPR & FPR & {\makecell{{AU-}\\{ROC}}} & TPR & FPR \\ %
					\cline{1-13}
					30 & 0.92 & 0.859 & 0.088 & 0.97 & 0.937 & 0.050 &  &  & & 0.92 & 0.784  & 0.048 \\ 
					\cline{1-7}\cline{11-13}
					60 & 0.94 &  0.829 & 0.029 & 0.97 & 0.934 & 0.039 & 0.91 & 0.799  & 0.071 & 0.94 & 0.822 & 0.043 \\   
					\cline{1-7}\cline{11-13}
					120 & 0.95 & 0.848 & 0.022 & 0.98 & 0.926 & 0.059 &  &  &  & 0.95 & 0.861 & 0.043 \\ 
					\hline
			\end{tabular}}
		\end{table*}


\begin{table*}[!bp]
	\centering
	\caption{Experiment 2 Results (NSL-KDD): TP, TN, FP, FN}
	\label{table:exp2cmnsl}
			{\begin{tabular}{|c|c|c|c|c|c|c|c|c|c|c|c|c|c|c|c|c|}%
					\cline{1-17}
					\multicolumn{1}{|c|}{\textbf{\# of}} & \multicolumn{16}{|c|}{Method} \\
					\cline{2-17}
					\multicolumn{1}{|c|}{\textbf{labelled}} & \multicolumn{4}{|c|}{\textbf{Weakly 2-sample}} & \multicolumn{4}{|c|}{\textbf{DevNet}} & \multicolumn{4}{|c|}{\textbf{Hybrid}} & \multicolumn{4}{|c|}{\textbf{Proposed}} \\%
					\cline{2-17}
					\textbf{anomalies} & TP & TN & FP & FN & TP & TN & FP & FN & TP & TN & FP & FN & TP & TN & FP & FN \\ %
					\cline{1-17}
					30 & 286 & 2725 & 264 & 47 & 312 & 2839 & 150 & 21 &  &  &  &  & 261 & 2845 & 144 & 72 \\ 
					\cline{1-9}\cline{14-17}
					60 & 276 & 2902 & 87 & 57 & 311 & 2871 & 118 & 22 & 266 & 2777 & 212 & 67 & 274 & 2861 & 128 & 59 \\  
					\cline{1-9}\cline{14-17}
					120 & 282 & 2925 & 64 & 51 & 308  & 2814 & 175 & 25 & & & & & 287 & 2862 & 127 & 46 \\
					\hline
			\end{tabular}}
		\end{table*}
		

\begin{table*}[!htbp]
	\centering
	\caption{Experiment 2 Results (CIC-IDS2018): AUROC, TPR, FPR}
	\label{table:exp2resids}
			{\begin{tabular}{|c|c|c|c|c|c|c|c|c|c|c|c|c|}%
					\cline{1-13}
					\multicolumn{1}{|c|}{\textbf{\# of}} & \multicolumn{12}{|c|}{Method} \\
					\cline{2-13}
					\multicolumn{1}{|c|}{\textbf{labelled}} & \multicolumn{3}{|c|}{\textbf{Weakly 2-sample}} & \multicolumn{3}{|c|}{\textbf{DevNet}} & \multicolumn{3}{|c|}{\textbf{Hybrid}} & \multicolumn{3}{|c|}{\textbf{Proposed}} \\%
					\cline{2-13}
					\textbf{anomalies} & {\makecell{{AU-}\\{ROC}}} & TPR & FPR & {\makecell{{AU-}\\{ROC}}} & TPR & FPR & {\makecell{{AU-}\\{ROC}}} & TPR & FPR & {\makecell{{AU-}\\{ROC}}} & TPR & FPR \\ %
					\cline{1-13}
					30 & 0.73 & 0.667 & 0.193 & 0.79 & 0.680 & 0.163 &  &  &  & 0.72 & 0.671 & 0.266\\ 
					\cline{1-7}\cline{11-13}
					60 & 0.75 & 0.714 & 0.366 & 0.80 & 0.716 & 0.168 & 0.69 & 0.683 & 0.261 & 0.76 & 0.666 & 0.318 \\   
					\cline{1-7}\cline{11-13}
					120 & 0.77 & 0.681 & 0.211 & 0.80 & 0.708 & 0.164 &  &  &  & 0.77 & 0.634 & 0.268 \\ 
					\hline					
			\end{tabular}}
		\end{table*}
			
	
	\begin{table*}[!htbp]
		\centering
		\caption{Experiment 2 Results (CIC-IDS2018): TP, TN, FP, FN}
		\label{table:exp2cmids}
				{\begin{tabular}{|c|c|c|c|c|c|c|c|c|c|c|c|c|c|c|c|c|}%
						\cline{1-17}
						\multicolumn{1}{|c|}{\textbf{\# of}} & \multicolumn{16}{|c|}{Method} \\
						\cline{2-17}
						\multicolumn{1}{|c|}{\textbf{labelled}} & \multicolumn{4}{|c|}{\textbf{Weakly 2-sample}} & \multicolumn{4}{|c|}{\textbf{DevNet}} & \multicolumn{4}{|c|}{\textbf{Hybrid}} & \multicolumn{4}{|c|}{\textbf{Proposed}} \\%
						\cline{2-17}
						\textbf{anomalies} & TP & TN & FP & FN & TP & TN & FP & FN & TP & TN & FP & FN & TP & TN & FP & FN \\ %
						\cline{1-17}
						30 & 331 & 3587 & 857 & 165 & 337 & 3721 & 723 & 159 &  &  &  &  & 333 & 3260 & 1184 & 163\\ 
						\cline{1-9}\cline{14-17}
						60 & 354 & 2818 & 1626 & 142 & 355 & 3699 & 745 & 141 & 339 & 3284 & 1160 & 157 & 330 & 3033 & 1411 & 166\\  
						\cline{1-9}\cline{14-17}
						120 & 338 & 3507 & 937 & 158 & 351 & 3716 & 728 & 145 &  &   &  &  & 315 & 3252 & 1192 & 181 \\
						\hline						
				\end{tabular}}
			\end{table*}

For CIC-IDS2018 dataset, as shown in Tables \ref{table:exp2resids} and \ref{table:exp2cmids}, AUROC for the proposed method increased by around 5\% when the number of labelled anomalies is increased to 60 from 30. Improvement in AUROC performance with increase in the number of labelled anomalies was similar in the case of weakly supervised 2-sample approach while it was modest for DevNet. However, the increase was around 1\% when the number of labelled anomalies was increased from 60 to 120 for the proposed method and similar improvement was observed across the other two methods.

For TON\_IoT dataset, improvement in AUROC during each step of number of anomalies is not drastic as the initial AUROC is already high as shown in Tables \ref{table:exp2reston} and \ref{table:exp2cmton}. It may be due to the less complex features or small number of samples present in the training and the test set. It is evident in high AUROC performance for all methods in this dataset. 

From the experimental results observed in Tables \ref{table:exp2resnsl}, \ref{table:exp2resids} and \ref{table:exp2reston}, the importance of using labelled anomaly samples for these datasets can be observed. Hybrid method is unable to benefit from labelled anomaly samples as it can not utilize them to improve its performance. On the other hand, rest of the methods mostly outperform the hybrid method with the help of additional labelled anomaly samples.


\begin{table*}[!bp]
	\centering
	\caption{Experiment 2 Results TON\_IoT (Win10): AUROC, TPR, FPR}
	\label{table:exp2reston}
			{\begin{tabular}{|c|c|c|c|c|c|c|c|c|c|c|c|c|}%
					\cline{1-13}
					\multicolumn{1}{|c|}{\textbf{\# of}} & \multicolumn{12}{|c|}{Method} \\
					\cline{2-13}
					\multicolumn{1}{|c|}{\textbf{labelled}} & \multicolumn{3}{|c|}{\textbf{Weakly 2-sample}} & \multicolumn{3}{|c|}{\textbf{DevNet}} & \multicolumn{3}{|c|}{\textbf{Hybrid}} & \multicolumn{3}{|c|}{\textbf{Proposed}} \\%
					\cline{2-13}
					\textbf{anomalies} & {\makecell{{AU-}\\{ROC}}} & TPR & FPR & {\makecell{{AU-}\\{ROC}}} & TPR & FPR & {\makecell{{AU-}\\{ROC}}} & TPR & FPR & {\makecell{{AU-}\\{ROC}}} & TPR & FPR \\ %
					\cline{1-13}
					30 & 0.98 & 0.960 & 0.047 & 0.99 & 0.979 & 0.001 &  &  & & 0.97 & 0.943 & 0.090 \\ 
					\cline{1-7}\cline{11-13}
					60 & 0.99 & 0.965 & 0.023 & 0.99 & 1.00 & 0.001 & 0.90 & 0.833  & 0.205 & 0.98 & 0.958 & 0.083 \\   
					\cline{1-7}\cline{11-13}
					120 & 0.99 & 0.968 & 0.022 & 0.99 & 1.00 & 0.001 &  &  &  & 0.99 & 0.981 & 0.008 \\ 
					\hline
			\end{tabular}}
		\end{table*}

		
		\begin{table*}[!bp]
			\centering
			\caption{Experiment 2 Results TON\_IoT (Win10): TP, TN, FP, FN}
			\label{table:exp2cmton}
					{\begin{tabular}{|c|c|c|c|c|c|c|c|c|c|c|c|c|c|c|c|c|}%
							\cline{1-17}
							\multicolumn{1}{|c|}{\textbf{\# of}} & \multicolumn{16}{|c|}{Method} \\
							\cline{2-17}
							\multicolumn{1}{|c|}{\textbf{labelled}} & \multicolumn{4}{|c|}{\textbf{Weakly 2-sample}} & \multicolumn{4}{|c|}{\textbf{DevNet}} & \multicolumn{4}{|c|}{\textbf{Hybrid}} & \multicolumn{4}{|c|}{\textbf{Proposed}} \\%
							\cline{2-17}
							\textbf{anomalies} & TP & TN & FP & FN & TP & TN & FP & FN & TP & TN & FP & FN & TP & TN & FP & FN \\ %
							\cline{1-17}
							30 & 46 & 413 & 21 & 2 & 47 & 433 & 1 & 1 &  &  &  &  & 45 & 395 & 39 & 3 \\ 
							\cline{1-9}\cline{14-17}
							60 & 46 & 424 & 10 & 2 & 48 & 433 & 1 & 0 & 40 & 345 & 89 & 8 & 46 & 398 & 36 & 2 \\  
							\cline{1-9}\cline{14-17}
							120 & 46 & 425 & 9 & 2 & 48 & 434 & 0 & 0 & & & & & 47 & 431 & 3 & 1 \\
							\hline
					\end{tabular}}
				\end{table*}
				
		
		\begin{table*}[!htbp]
			\centering
			\caption{Experiment 3 Results (NSL-KDD): AUROC, TPR, FPR}
			\label{table:exp3resnsl}
					{\begin{tabular}{|c|c|c|c|c|c|c|c|c|c|c|c|c|}%
							\cline{1-13}
							\multicolumn{1}{|c|}{} & \multicolumn{12}{|c|}{Method} \\
							\cline{2-13}
							\multicolumn{1}{|c|}{\textbf{Anomaly}} & \multicolumn{3}{|c|}{\textbf{Weakly 2-sample}} & \multicolumn{3}{|c|}{\textbf{DevNet}} & \multicolumn{3}{|c|}{\textbf{Hybrid}} & \multicolumn{3}{|c|}{\textbf{Proposed}} \\%
							\cline{2-13}
							\textbf{Percent} & {\makecell{{AU-}\\{ROC}}} & TPR & FPR & {\makecell{{AU-}\\{ROC}}} & TPR & FPR & {\makecell{{AU-}\\{ROC}}} & TPR & FPR & {\makecell{{AU-}\\{ROC}}} & TPR & FPR \\ %
							\cline{1-13}
							2 & 0.97 & 0.927 & 0.062 & 0.98 & 0.959 & 0.044 & 0.91 & 0.841 & 0.151 & 0.97 & 0.910 & 0.062\\ 
							\cline{1-13}
							5 & 0.96 & 0.880 & 0.035 & 0.98 & 0.944 & 0.035 & 0.90 & 0.874 & 0.166 & 0.95 & 0.862 & 0.049\\   
							\cline{1-13}
							10 & 0.94 & 0.829 & 0.029 & 0.97 & 0.934 & 0.039 & 0.91 & 0.799 & 0.071 & 0.94 & 0.822 & 0.043 \\ 
							\hline					
					\end{tabular}}
				\end{table*}
						
			
	\begin{table*}[!htbp]
		\centering
		\caption{Experiment 3 Results (NSL-KDD): TP, TN, FP, FN}
		\label{table:exp3cmnsl}
				{\begin{tabular}{|c|c|c|c|c|c|c|c|c|c|c|c|c|c|c|c|c|}%
						\cline{1-17}
						\multicolumn{1}{|c|}{} & \multicolumn{16}{|c|}{Method} \\
						\cline{2-17}
						\multicolumn{1}{|c|}{\textbf{Anomaly}} & \multicolumn{4}{|c|}{\textbf{Weakly 2-sample}} & \multicolumn{4}{|c|}{\textbf{DevNet}} & \multicolumn{4}{|c|}{\textbf{Hybrid}} & \multicolumn{4}{|c|}{\textbf{Proposed}} \\%
						\cline{2-17}
						\textbf{Percent} & TP & TN & FP & FN & TP & TN & FP & FN & TP & TN & FP & FN & TP & TN & FP & FN \\ %
						\cline{1-17}
						2 & 309 & 2805 & 184 & 24 & 320 & 2858 & 131 & 13 & 280 & 2538 & 451 & 53 & 303 & 2804 & 185 & 30\\ 
						\cline{1-17}
						5 & 293 & 2884 & 105 & 40 & 314 & 2884 & 105 & 19 & 291 & 2493 & 496 & 42 & 287 & 2843 & 145 & 46\\  
						\cline{1-17}
						10 & 276 & 2902 & 87 & 57 & 311 & 2871 & 118 & 22 & 266 & 2777 & 212 & 67 & 274 & 2593 & 128 & 59 \\
						\hline						
				\end{tabular}}
			\end{table*}


\begin{table*}[!bp]
	\centering
	\caption{Experiment 3 Results (CIC-IDS2018): AUROC, TPR, FPR}
	\label{table:exp3resids}
			{\begin{tabular}{|c|c|c|c|c|c|c|c|c|c|c|c|c|}%
					\cline{1-13}
					\multicolumn{1}{|c|}{} & \multicolumn{12}{|c|}{Method} \\
					\cline{2-13}
					\multicolumn{1}{|c|}{\textbf{Anomaly}} & \multicolumn{3}{|c|}{\textbf{Weakly 2-sample}} & \multicolumn{3}{|c|}{\textbf{DevNet}} & \multicolumn{3}{|c|}{\textbf{Hybrid}} & \multicolumn{3}{|c|}{\textbf{Proposed}} \\%
					\cline{2-13}
					\textbf{Percent} & {\makecell{{AU-}\\{ROC}}} & TPR & FPR & {\makecell{{AU-}\\{ROC}}} & TPR & FPR & {\makecell{{AU-}\\{ROC}}} & TPR & FPR & {\makecell{{AU-}\\{ROC}}} & TPR & FPR \\ %
					\cline{1-13}
					2 & 0.80 & 0.724 & 0.210 & 0.81 & 0.724 & 0.201 & 0.71 & 0.561 & 0.243 & 0.78 & 0.673 & 0.227 \\ 
					\cline{1-13}
					5 & 0.79 & 0.719 & 0.232 & 0.80 & 0.705 & 0.183 & 0.70 & 0.673 & 0.282 & 0.78 & 0.684 & 0.290 \\   
					\cline{1-13}
					10 & 0.75 & 0.714 & 0.366 & 0.80 & 0.716 & 0.168 & 0.69 & 0.683 & 0.261 & 0.76 & 0.666 & 0.318 \\ 
					\hline
			\end{tabular}}
		\end{table*}

	
	\begin{table*}[!htbp]
		\centering
		\caption{Experiment 3 Results (CIC-IDS2018): TP, TN, FP, FN}
		\label{table:exp3cmids}
				{\begin{tabular}{|c|c|c|c|c|c|c|c|c|c|c|c|c|c|c|c|c|}%
						\cline{1-17}
						\multicolumn{1}{|c|}{} & \multicolumn{16}{|c|}{Method} \\
						\cline{2-17}
						\multicolumn{1}{|c|}{\textbf{Anomaly}} & \multicolumn{4}{|c|}{\textbf{Weakly 2-sample}} & \multicolumn{4}{|c|}{\textbf{DevNet}} & \multicolumn{4}{|c|}{\textbf{Hybrid}} & \multicolumn{4}{|c|}{\textbf{Proposed}} \\%
						\cline{2-17}
						\textbf{Percent} & TP & TN & FP & FN & TP & TN & FP & FN & TP & TN & FP & FN & TP & TN & FP & FN \\ %
						\cline{1-17}
						2 & 359 & 3509 & 935 & 137 & 359 & 3551 & 893 & 137 & 278 & 3364 & 1080 & 218 & 334 & 3436 & 1008 & 162 \\ 
						\cline{1-17}
						5 & 357 & 3414 & 1030 & 139 & 350 & 3630 & 814 & 146 & 334 & 3191 & 1253 & 162 & 339 & 3157 & 1287 & 157 \\  
						\cline{1-17}
						10 & 354 & 2818 & 1626 & 142 & 355  & 3699 & 745 & 141 & 339 & 3284 & 1160 & 157 & 330 & 3033 & 1411 & 166 \\
						\hline
				\end{tabular}}
			\end{table*}
				
		
		\begin{table*}[!htbp]
			\centering
			\caption{Experiment 3 Results TON\_IoT (Win10): AUROC, TPR, FPR}
			\label{table:exp3reston}
					{\begin{tabular}{|c|c|c|c|c|c|c|c|c|c|c|c|c|}%
							\cline{1-13}
							\multicolumn{1}{|c|}{} & \multicolumn{12}{|c|}{Method} \\
							\cline{2-13}
							\multicolumn{1}{|c|}{\textbf{Anomaly}} & \multicolumn{3}{|c|}{\textbf{Weakly 2-sample}} & \multicolumn{3}{|c|}{\textbf{DevNet}} & \multicolumn{3}{|c|}{\textbf{Hybrid}} & \multicolumn{3}{|c|}{\textbf{Proposed}} \\%
							\cline{2-13}
							\textbf{Percent} & {\makecell{{AU-}\\{ROC}}} & TPR & FPR & {\makecell{{AU-}\\{ROC}}} & TPR & FPR & {\makecell{{AU-}\\{ROC}}} & TPR & FPR & {\makecell{{AU-}\\{ROC}}} & TPR & FPR \\ %
							\cline{1-13}
							2 & 0.99 & 1.00 & 0.002 & 0.99 & 1.00 & 0.001 & 0.93 & 0.943 & 0.162 & 0.99 & 0.997 & 0.004\\ 
							\cline{1-13}
							5 & 0.99 & 0.988 & 0.009 & 0.99 & 1.00 & 0.001 & 0.91 & 0.867 & 0.145 & 0.99 & 0.994 & 0.003 \\   
							\cline{1-13}
							10 & 0.99 & 0.965 & 0.023 & 0.99 & 1.00 & 0.001 & 0.90 & 0.833 & 0.205 & 0.98 & 0.958 & 0.083 \\ 
							\hline					
					\end{tabular}}
				\end{table*}
				
		
		\begin{table*}[!htbp]
			\centering
			\caption{Experiment 3 Results TON\_IoT (Win10): TP, TN, FP, FN}
			\label{table:exp3cmton}
					{\begin{tabular}{|c|c|c|c|c|c|c|c|c|c|c|c|c|c|c|c|c|}%
							\cline{1-17}
							\multicolumn{1}{|c|}{} & \multicolumn{16}{|c|}{Method} \\
							\cline{2-17}
							\multicolumn{1}{|c|}{\textbf{Anomaly}} & \multicolumn{4}{|c|}{\textbf{Weakly 2-sample}} & \multicolumn{4}{|c|}{\textbf{DevNet}} & \multicolumn{4}{|c|}{\textbf{Hybrid}} & \multicolumn{4}{|c|}{\textbf{Proposed}} \\%
							\cline{2-17}
							\textbf{Percent} & TP & TN & FP & FN & TP & TN & FP & FN & TP & TN & FP & FN & TP & TN & FP & FN \\ %
							\cline{1-17}
							2 & 48 & 433 & 1 & 0 & 48 & 433 & 1 & 0 & 45 & 364 & 70 & 3 & 48 & 432 & 2 & 0\\ 
							\cline{1-17}
							5 & 47 & 430 & 4 & 1 & 48 & 434 & 0 & 0 & 42 & 371 & 63 & 6 & 47 & 433 & 1 & 1\\  
							\cline{1-17}
							10 & 46 & 424 & 10 & 2 & 48 & 433 & 1 & 0 & 40 & 345 & 89 & 8 & 46 & 398 & 36 & 2 \\
							\hline						
					\end{tabular}}
				\end{table*}

\subsection{Experiment 3: Varying the anomaly percentage}

To evaluate the impact of anomaly percentage on the overall anomaly detection, we fixed the number of anomalies at 60, while varying the anomaly percentage. For this experiment, the anomaly percentage was varied between $[2,5,10]$ for each dataset.

For NSL-KDD dataset, interestingly, while other methods suffer a slight decrease in the AUROC performance with increase in anomaly percentage, the hybrid method reports relatively unchanged AUROC performance as shown in Tables \ref{table:exp3resnsl} and \ref{table:exp3cmnsl}. The proposed method displays similar TPR and FPR compared to the other methods for different anomaly percentages.

For CIC-IDS2018 dataset, a lower AUROC is observed across all the methods compared to other datasets as shown in Tables \ref{table:exp3resids} and \ref{table:exp3cmids}. TPR for the proposed method is generally better than Hybrid method, whereas the DevNet generally has a better overall performance.

For TON\_IoT dataset, as shown in Tables \ref{table:exp3reston} and \ref{table:exp3cmton}, the AUROC performance is high for all methods irrespective of changes in anomaly percentage. The proposed method generally shows high TPR and low FPR similar to other methods using the labelled anomaly set.

As evident from this experiment, hybrid method without any labelled anomaly samples displays performance trends similar to remaining methods which utilize the labelled anomaly samples. As the number of labelled anomaly samples are kept constant during the course of this experiment, methods utilizing these samples mostly outperform the hybrid method owing to its higher baseline performance despite showing similar trends with increasing anomaly percentage.


\section{Discussion}\label{sec:discussion}

For all three datasets, generally the AUROC performance improved with the increase in the number of labelled anomalies as shown in Tables \ref{table:exp2resnsl} to \ref{table:exp2cmton}. Larger improvement in performance was observed when the number of labelled anomalies was increased from 30 to 60, compared to the improvement when the change was from 60 to 120.

For all three datasets, generally the AUROC performance decreases with the increase in the anomaly percentage as shown in Tables \ref{table:exp3resnsl} to \ref{table:exp3cmton}. However, the decrease in the performance does not appear to be significant for any of the methods. This observation suggests that all the methods under consideration, including the proposed method, are robust to some changes in anomaly percentage.

For CIC-IDS2018 dataset in particular, a lower AUROC is observed compared to other datasets as shown in Tables \ref{table:exp2resids}, \ref{table:exp2cmids}, \ref{table:exp3resids} and \ref{table:exp3cmids}. This is consistent across all the evaluation methods which may indicate difficulty in the separation between normal and anomalous samples for this dataset.

For the datasets considered in this paper, benefits of using limited labelled anomaly samples was observed through the experiments 2 and 3 as hybrid method was generally outperformed by rest of the methods.
	
To further improve the performance of the proposed few-shot weakly supervised method, deeper machine learning algorithms can be explored for the anomaly scoring stage depending on the complexity of the datasets. Deeper networks could be particularly useful for datasets such as CIC-IDS2018 where the performance of the existing methods has a scope for improvements. Additionally, instead of only 3 samples, more labelled anomaly samples can be combined to form a single sample of the augmented dataset which would be able to generate larger useful combinations of samples. Both these approaches can be combined to identify the suitable approach depending on the complexity of dataset.


\section{Conclusion}\label{sec:conclusion}
We proposed an enhanced three-stage few-shot weakly supervised deep learning anomaly detection framework. The data augmentation stage created additional training samples by combining three samples from the unlabelled and limited labelled anomaly set. Anomaly scoring network and ordinal regression stages were combined to generate corresponding anomaly scores for each sample. We evaluated the performance on three datasets: NSL-KDD, CIC-IDS2018, and TON\_IOT (Windows 10) to achieve comparable anomaly detection performance to the state-of-the-art methods. Utilization of deeper machine learning algorithms in the anomaly scoring stage of the framework can be explored as future extension of this work.


\bibliographystyle{elsarticle-num}
\bibliography{main.bib}

\end{document}